\documentclass[aps,prb,twocolumn,floatfix,superscriptaddress,showpacs,reprint,longbilbiography]{revtex4-2}
\usepackage{amsmath, amssymb, amsthm, amsfonts}
\usepackage{bm} 
\usepackage{graphicx, subfigure}
\usepackage{color}
\usepackage{times}
\usepackage{xfrac}
\usepackage[pdftex,colorlinks]{hyperref}
\hypersetup{
    colorlinks=true,
    linkcolor=blue,
    filecolor=blue,
    urlcolor=blue,
    citecolor=blue,
}
\newcommand{\Rnum}[1]{\uppercase\expandafter{\romannumeral#1}}
\newcommand{\bs}[1]{\boldsymbol{#1}}

\begin{document}

\title{Universal scaling behavior of resistivity under two-dimensional superconducting phase fluctuations}
\author{Zongsheng Zhou}
\affiliation{Beijing National Laboratory for Condensed Matter Physics and Institute of Physics, Chinese Academy of Sciences, Beijing 100190, China}

\author{Kang Wang}
\affiliation{Beijing National Laboratory for Condensed Matter Physics and Institute of Physics, Chinese Academy of Sciences, Beijing 100190, China}
\affiliation{School of Physical Sciences, University of Chinese Academy of Sciences, Beijing 100049, China}

\author{Hai-Jun Liao}
\affiliation{Beijing National Laboratory for Condensed Matter Physics and Institute of Physics, Chinese Academy of Sciences, Beijing 100190, China}
\affiliation{Songshan Lake Materials Laboratory, Dongguan, Guangdong 523808, China}

\author{Zi-Xiang Li}
\email{zixiangli@iphy.ac.cn}
\affiliation{Beijing National Laboratory for Condensed Matter Physics and Institute of Physics, Chinese Academy of Sciences, Beijing 100190, China}
\affiliation{School of Physical Sciences, University of Chinese Academy of Sciences, Beijing 100049, China}

\author{Tao Xiang}
\email{txiang@iphy.ac.cn}
\affiliation{Beijing National Laboratory for Condensed Matter Physics and Institute of Physics, Chinese Academy of Sciences, Beijing 100190, China}
\affiliation{School of Physical Sciences, University of Chinese Academy of Sciences, Beijing 100049, China}
\affiliation{Beijing Academy of Quantum Information Sciences, Beijing, 100190, China}

\date{\today}

\begin{abstract}
    In superconductors with relatively low superfluid density, such as cuprate high-$T_c$ superconductors, the phase fluctuations of the superconducting order parameter are remarkable, presumably playing a nonnegligible role in shaping many distinctive physical properties. This work systematically investigates the electrical transport properties arising from thermal superconducting phase fluctuations in two-dimensional superconductors. Employing the Monte Carlo procedure, we access the numerically exact properties of a microscopic model of superconductivity, in which the classical XY model governs the thermal phase fluctuations of the superconducting order parameter.  For both $s$-wave and $d_{x^2-y^2}$-wave pairings, the electrical resistivity exhibits a universal scaling behavior in the temperature range above $T_c$. Our numerical results demonstrate that the scaling behavior of the quasiparticle lifetime is associated with the correlation length of the superconducting order parameter, yielding the universal scaling behavior of electrical resistivity determined by the Berezinskii-Kosterlitz-Thouless critical scaling of the correlation length. Furthermore, we discuss the dependence of the electrical resistivity coefficient on the pairing amplitude and the possible implication on recent transport experiments. 
\end{abstract}

\maketitle

\section{Introduction}
 Superconductivity (SC) is a macroscopic quantum phenomenon stemming from the condensation of paired electrons, which is characterized by a complex pairing order parameter with both amplitude and phase~\cite{XiangT2022CUP, Combescot2022CUP}. For BCS mean-field theory, it is widely assumed that the phase fluctuations of the order parameter are negligible for determining important physical properties such as transition temperature $T_c$, which generically holds for SC in many conventional metals~\cite{Bardeen1957PRB}. However, phase fluctuations in conventional superconductors could be enlarged by disorder~\cite{Mondal2011PRL, Pratap2022JPCM, Weitzel2023PRL}. Additionally, for unconventional superconductors, including cuprates~\cite{Uemura2004JPCM, Hetel2007NP, Keimer2015Nat, Bozovic2016Nat}, organic superconductors~\cite{Ardavan2012JPSJ, Jotzu2023PRX}, and twisted bilayer graphene\cite{Hazrz2019PRX, HuXiang2019PRL}, the superfluid density is low owing to their low charge carrier density, giving rise to the strong phase fluctuations that presumably accounts for many intriguing physical phenomena~\cite{Emery1995Nat}. In the past decades, phase fluctuations in unconventional superconductors have been extensively investigated from both theoretical~\cite{Randeria1992PRL, Franz1998PRB, EcklT2002PRB, Mayr2005PRL, Larkin2005OUP, EcklT2006PRB, ErezBerg2007PRL, HQiang2010PRB, ZhongYongwei2011PRB, Litao2011JPCM, LiaoHJ2011JPCM, Groshev2020JETP, LiZX2021npjQM, Singh2022PRB, WangXucheng2023PRB, Harun2023SPPC, LiZX2023PRB} and experimental~\cite{XuZA2000Nat, WangYY2005PRL, Hetel2007NP, Hufner2008RPP, LiLu2010PRB, Rourke2011NP, Chang2012NP, XiaoH2014PRB, XiaoH2016JPCM, Matsuda2016NC, Faeth2021PRX, Chenheyu2021PRX, ChenSD2022Nat} perspectives. In the presence of strong phase fluctuations, the phase coherence temperature is lower than the temperature at which the Copper pairing forms, leading to the opening of the spectral gap above superconducting $T_c$~\cite{Randeria1992PRL, Emery1995Nat, Franz1998PRB, EcklT2002PRB, EcklT2006PRB, ErezBerg2007PRL, HQiang2010PRB, WangXucheng2023PRB}. It is also theoretically proposed that phase fluctuations could trigger novel phenomena such as the Fermi arc~\cite{ErezBerg2007PRL, HQiang2010PRB, ZhongYongwei2011PRB, Litao2011JPCM, WangXC2013Arxiv} observed in underdoped cuprates~\cite{Marshall1996PRL, DingH1996Nat, Loeser1996Sci}, and the possibility of charge-4e SC~\cite{ErezBerg2009NP, JiangYF2017PRB, JianSK2021PRL, Genzdilov2022PRB, LiPF2022Arxiv, QinQ2023PRB, WuYM2023Arxiv, GeJun2024PRX} as a vestigial order.  

 Nevertheless, the transport properties of unconventional superconductors, including electrical resistivity arising from superconducting phase fluctuations in the temperature regime above $T_c$, are rarely studied. Recently, armed with the development of thin-film technology, strange-metal electrical transport behaviors, which were unveiled in underdoped and optimally doped cuprates soon after the discovery of high-$T_c$ superconductors~\cite{Fiory1987PRL, Martin1990PRB, Takagi1992PRL}, have been observed in a wide doping range of electron-doped cuprates $\rm La_{2-x}Ce_xCuO_4$~\cite{JinK2011Nat, YuanJie2022Nat} and iron-pnictide superconductors $\rm FeSe$~\cite{JiangXY2023NP}, along with ${\rm Ca}_{10}\left({\rm Pt}_4 {\rm As}_8\right)\left(\left({\rm Fe}_{0.97}{\rm Pt}_{0.03}\right)_2{\rm As}_2\right)_5$~\cite{CaiShu2023NC}. More remarkably, the coefficient of linear-$T$ resistivity exhibits a quadratic dependence on the superconducting transition temperature $T_c$~\cite{YuanJie2022Nat, JiangXY2023NP, CaiShu2023NC}, which implies the possible relation between anomalous metallic transport and the SC or other ingredients associated with superconducting pairing mechanisms. Hence, exploring the electrical resistivity driven by the superconducting phase fluctuations above $T_c$ is desirable, which potentially sheds new insights into understanding the possible underlying relation between strange metal dissipation and SC observed in unconventional superconductors.  

 In this work, we systematically investigate the properties of electrical transports of a microscopic BdG Hamiltonian involving superconducting pairing and thermal phase fluctuations above the superconducting critical temperature $T_c$. The amplitude of the superconducting order parameter is fixed as a spatially independent value.  Distinct from the BCS mean-field Hamiltonian, the model under consideration incorporates the spatial fluctuations of phase. The phase fluctuations are governed by a classical XY model, characterizing the thermal fluctuations of the order parameter's phase with varying temperatures. Employing large-scale Monte Carlo simulation, we study the transport properties of the model at finite temperatures. The numerical results reveal that the electrical resistivity induced by the thermal phase fluctuations in the low-temperature regime above $T_c$ obeys the scaling behavior $\rho\left(T\right) \propto A \mathrm{exp}({-b}/{\sqrt{T/T_c-1}})$, which is consistent with the critical scaling relation of the correlation length for the Berezinskii-Kosterlitz-Thouless (BKT) transition~\cite{Kosterlitz1974JPC}. We further evaluate the quasiparticle lifetime of electrons versus temperature. Interestingly, we achieve the same scaling behavior as that of electrical resistivity. Hence, our numerical results establish that thermal phase fluctuations yield the anomalous temperature dependence of electrical resistivity in two-dimensional superconductors, which is determined by the BKT critical scaling of the correlation length. 

 This paper is organized as follows: Section~\Rnum{2} introduces the model Hamiltonian, which leverages the classical two-dimensional XY model to simulate the superconducting phase fluctuations, along with the approach for determining resistance values through finite lattice scaling analysis. Section~\Rnum{3} presents the main results. Firstly, we elucidate the temperature dependence of DC resistivity, revealing universal BKT behavior, and show numerical evidence for a quadratic scaling relationship between the coefficient of temperature-dependent electrical resistivity and the superconducting pairing amplitude. Secondly, we focus on determining the superconducting transition temperature $T_c$ from the superfluid density.  The paper concludes with a summary of the key findings in Section~\Rnum{4}.

\section{Model}

 In superconductors characterized by pronounced superconducting fluctuations, such as cuprates, the superconducting transition temperature $T_c$ is often significantly lower than the transition temperature $T_{\rm MF}$ predicted by the mean-field theory. Within the temperature regime above $T_c$ yet substantially below $T_{\rm MF}$, superconducting coherence is compromised due to phase fluctuations, whose effects can be effectively modeled by the BdG Hamiltonian:
\begin{equation}
    \mathcal{H}_0 \! =\!  \sum_{\bs{r},\bs{r}^\prime}
    \begin{pmatrix}
        c_{\bs{r}\uparrow}^\dagger ,  c_{\bs{r}\downarrow} 
    \end{pmatrix}\!
    \begin{pmatrix}
        H_0 (\bs{r},\bs{r^\prime} ) & \Delta (\bs{r}, \bs{r}^\prime )\\
        \Delta^* (\bs{r},\bs{r}^\prime ) & H_0 (\bs{r}, \bs{r}^\prime )
    \end{pmatrix} \!
    \begin{pmatrix}
        c_{\bs{r}^\prime\uparrow} \\
        c_{\bs{r}^\prime\downarrow}^{\dagger}
    \end{pmatrix} ,
\label{Ham}
\end{equation}
 where $c_{\bs{r}\sigma}^\dagger $ is the creation operator of an electron at site $\bs{r}$ with spin $\sigma$ and $H_0$ is the Hamiltonian of non-interacting electrons defined on a square lattice, 
\begin{equation}
    H_0 (\bs{r}, \bs{r}^\prime )= - t \delta_{\bs{r}^\prime,\bs{r}\pm\bs{e}}-\mu\delta_{\bs{r},\bs{r}^\prime},
\end{equation}
 with $t$ the hopping integral between two nearest-neighboring sites,  $\bs{e}=(\bs{e}_x, \bs{e}_y)$ the two unit lattice vectors, and $\mu$ the chemical potential. Variable $\theta$ is the phase of the superconducting gap parameter $\Delta (\bs{r}, \bs{r}^\prime )$  
\begin{equation}
 \Delta (\bs{r}, \bs{r}^\prime ) = \left| \Delta (\bs{r}, \bs{r}^\prime )  \right| e^{i\theta (\bar{\bs{r}})},
\end{equation}
 where $\bar{\bs{r}}= (\bs{r}+\bs{r}^\prime)/2$. In the temperature regime $T_c<T\ll T_{\rm MF}$, we assume the gap amplitude to be a constant, $\left| \Delta (\bs{r}, \bs{r}^\prime )  \right| = \Delta$, independent of $(\bs{r}, \bs{r}^\prime )$ and $T$.

 In this work, two kinds of pairing states are considered. One is the $s$-wave pairing state, and the other is the $d_{x^2-y^2}$-wave pairing state. The $s$-wave pairing state only involves the on-site pairing, whose gap parameter is defined by
\begin{equation}
	\Delta (\bs{r}, \bs{r}^\prime ) = \Delta e^{i\theta\left(\bs{r}\right)} \delta_{\bs{r}, \bs{r}^\prime }.
\end{equation}
 The $d_{x^2-y^2}$-wave pairing state, on the other hand, arises from the pairing of two electrons on two nearest-neighboring sites $\bs{r}$ and $\bs{r}^\prime=\bs{r}\pm\bs{e}$ and the phase parameter is defined on the corresponding link of the square lattice. 
\begin{equation}
   \Delta ( \bs{r}, \bs{r}^\prime ) = \Delta e^{i\theta (\bar{\bs{r}})}  
   \left( \delta_{{\bs r}^\prime , {\bs r}\pm {\bs e}_x } - \delta_{{\bs r}^\prime , {\bs r}\pm {\bs e}_y } \right). 
\end{equation}

 In the temperature range considered here, we assume that the fluctuation of $\theta$ is slow in comparison with the quasiparticle excitations so that the dynamics of $\theta$ are purely governed by the classical XY model, described by the Hamiltonian
\begin{equation}
    \mathcal{H}_\theta  =  - \sum_{\left\langle\bar{\bs{r}},\bar{\bs{r}}^\prime\right\rangle}J_{\theta}\mathrm{cos}
               \left[\theta\left(\bar{\bs{r}}\right) - \theta\left(\bar{\bs{r}}^\prime\right)\right] ,
\label{Ham_theta}
\end{equation}
 where $\langle \cdots \rangle$ denotes summations over the nearest-neighboring sites for the $s$-wave pairing state or bond centers for the $d_{x^2-y^2}$-wave pairing state.  The employment of the classical Monte Carlo method allows for the generation of a series of configurations of $\theta$ using this model at a given temperature. To mitigate self-correlation, a single configuration is selected from every 20,000 configurations produced by the Monte Carlo simulation. By inserting each of these phase configurations into Eq.~(\ref{Ham}), it becomes possible to determine the eigenvalues $\varepsilon_a$ and their corresponding eigenvectors $\psi_a$ of ${\cal H}_0$ by exact diagonalization.

 In the numerical simulations, we set the hopping integral $t = 1$ as the unit of energy, the coupling constant $J_\theta = 0.1$, the gap amplitude $\Delta = 0.3$, and the electron filling factor $n_e = 0.83$ (the filling factor is 2 if a band is fully filled) if not explicitly stated. We take these values in our numerical simulations such that there is a relatively wide temperature window above the superconducting transition but lower than the energy scale of pairing amplitude. Furthermore, periodic boundary conditions are assumed.

\section{Electromagnetic response functions}

 In the linear response theory, the electromagnetic response function is determined by the current-current correlation function
\begin{equation}
	\Lambda_{\alpha\beta}  (\bs{q},\omega ) = \int_0^{\infty} d t e^{i\omega t} \left[ j^\alpha (\bs{q}, t ), j^\beta (-\bs{q}, 0) \right],
\end{equation}
 where $j^\alpha (\bs{q}, t )$ is the Fourier component of the current operator along the $\alpha$-direction 
 \begin{equation}
    j^\alpha (\bs{r} ) = i t \sum_{\sigma} \left( c_{\bs{r},\sigma}^\dagger c_{\bs{r}+\bs{e}_\alpha,\sigma} - c_{\bs{r}+\bs{e}_\alpha ,\sigma}^\dagger c_{\bs{r},\sigma} \right).
\end{equation}
 
 Using the eigenpairs of ${\cal H}_0$, the current-current correlation can be expressed as
\begin{equation}
    \Lambda^{\alpha\beta} (\bs{q}, \omega ) = \frac{1}{N}\sum_{a\neq b}M _{ab} (\bs{q} ) M_{ba} (-\bs{q} )  \frac{f (\varepsilon_a ) - f (\varepsilon_b ) }{\omega -\varepsilon_a + \varepsilon_b + i\eta},\label{Lamab}
\end{equation}
where $N= L\times L$ is the lattice size, $f\left(\varepsilon\right)$ is the Fermi distribution function, $\eta$ is a positive number that tends to zero in the thermodynamic limit $N\rightarrow \infty$, and
\begin{eqnarray}
M_{ab} \left(\bs{q}\right) &=& \sum_{\bs{r}\sigma}\left[\psi_{a}^* (\bs{r},\sigma ) \psi_{b} (\bs{r}+\bs{e}_\alpha,\sigma)\right. \nonumber \\
&&-\left. \psi_{a}^* (\bs{r}+\bs{e}_\alpha,\sigma ) \psi_{b} (\bs{r},\sigma ) \right]e^{i\bs{q}\cdot \bs{r}}.
\end{eqnarray}

\subsection{Resistivity} 
\label{sec:resistivity}

\begin{figure}[t]
   \includegraphics{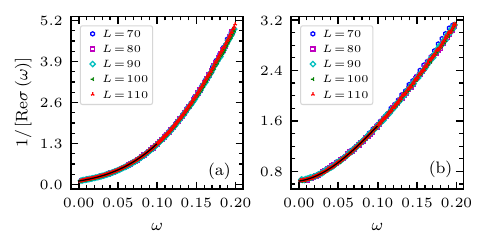}
   \caption{Frequency dependence of the inverse of $\mathrm{Re} \sigma (\omega )$ obtained on five different $L\times L$ square lattices at $T=0.1$ for the two phase fluctuating states with (a) $s$-wave and (b) $d_{x^2-y^2}$-wave pairing states.}
   \label{Fig1}
\end{figure}

 The resistivity is the reciprocal of the real part of optical conductivity $\sigma\left(\omega\right)$ in the $\omega\rightarrow 0$ limit. The optical conductivity, on the other hand, is determined by the imaginary part of the current-current correlation function in the long-wavelength limit 
\begin{eqnarray}
    \mathrm{Re}\sigma\left(\omega\right) = \frac{\mathrm{Im}\Lambda^{xx}\left(\bs{q}=\bs{0},\omega\right)}{\omega} .
  \label{Sig}
\end{eqnarray}
 Using the expression (\ref{Lamab}), it is straightforward to show that 
\begin{equation}
    \mathrm{Re}\sigma\left(\omega\right) = \sum_{a\neq b} W_{ab} \delta (\omega -\varepsilon_a +\varepsilon_b) 
    \label{Eq:sigma_finite}
\end{equation}
in the limit $\eta \rightarrow 0$. In this equation, 
\begin{equation}
    W_{ab} =- \frac{\pi}{N} M_{ab} (\bs{0}) M_{ba} (\bs{0} ) \frac{f (\varepsilon_a ) - f (\varepsilon_b)}{\varepsilon_a - \varepsilon_b} .
\end{equation}
 Hence, only the terms satisfying $\omega = \varepsilon_a - \varepsilon_b$ contribute to the conductivity.

 In a finite lattice system, the eigenvalues of ${\cal H}_0$ are discrete. Therefore, $\sigma (\omega)$ determined using Eq.~(\ref{Eq:sigma_finite}) is also discrete. To extrapolate the value of the conductivity to the thermodynamic limit, we first calculate the total weight of the conductivity for the frequency from 0 to $\omega$
 \begin{equation}
     S_\mathrm{tot} (\omega) = \sum_{\substack{a\not= b\\ {\footnotesize{ 0 < \varepsilon_a - \varepsilon_b <\omega} }}} W_{ab}. 
 \end{equation}
 We then determine the conductivity from the numerical derivative of $S_\mathrm{tot}$
 \begin{equation}
     \sigma (\omega)  = \frac{ S_\mathrm{tot} (\omega +d\omega) -  S_\mathrm{tot} (\omega - d\omega)}{2d\omega} ,
     \label{Eq:sigma}
 \end{equation} 
 where $d\omega$ is a small parameter that should be taken such that at least one eigenvalue difference $\varepsilon_a - \varepsilon_b$ falls between the interval ($\omega - d\omega, \omega + d\omega$). Furthermore, we need to do an average of $\sigma (\omega)$ over different phase configurations generated from the Monte-Carlo sampling of the XY model at a given temperature. The resistivity is finally obtained by extrapolating the inverse of the conductivity $\sigma (\omega)$ to the limit $\omega \rightarrow 0$. 
 
 Fig.~\ref{Fig1} is an illustrative example, demonstrating the variation in optical conductivity with frequency $\omega$ at $T=0.1$ according to the scheme above. Notably, the outcomes across various system sizes converge onto a single curve for both $s$- and $d_{x^2-y^2}$-wave pairing states. This convergence suggests that the scheme is highly effective in reducing the impact of finite-size effects, thereby enabling accurate extrapolation of resistivity.  

\begin{figure}[t]
    \includegraphics{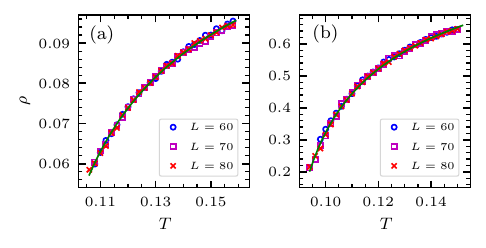}
    \caption{Temperature dependence of the DC resistivity for the $s$-wave (left panel)  and $d_{x^2-y^2}$-wave (right panel) pairing states in three different square lattice systems. The numerical results are well described by the formula $\rho\left(T\right) = A \mathrm{exp}({-b}/{\sqrt{T/T_{c}-1}})$, depicted by the solid curves. From the fitting, the superconducting transition temperature $T_c$ is found to be {$T_c=0.089(2)$} in both cases.}\label{Fig2}
\end{figure}

 Resistivity is determined by extrapolating $1/\mathrm{Re} \sigma (\omega)$ to the $\omega=0$ limit. Fig.~\ref{Fig2} depicts how resistivity varies with temperature for both $s$- and $d_{x^2-y^2}$-wave pairing states. This analysis is based on an average of over 72 independent phase configurations for each system size at a specific temperature. The findings reveal that resistivity stabilizes as system size increases, indicating negligible finite-size effects. Although there are quantitative differences in their absolute values, a universal temperature dependence emerges for both pairing states. Specifically, resistivity escalates with temperature, displaying a downward curvature that contrasts with the $T^2$ behavior characteristic of Fermi liquids.

 To delve deeper into the temperature-dependent behavior of resistivity, we conduct a scaling analysis on the numerical data. In the system under study, resistivity results from electron scattering by the fluctuating phase field. Given that the correlation length of the fluctuating pairing phase constitutes the sole length scale in the XY model, we propose that the electron scattering length $l_s$, which is a product of the scattering time $\tau_s$ and the Fermi velocity, is proportional to the correlation length $\xi$ of fluctuating phase, i.e., $l_s \propto \xi$. As the temperature approaches the BKT critical temperature $T_c$, the correlation length $\xi$ is expected to diverge, following the theory outlined by Kosterlitz in 1974~\cite{Kosterlitz1974JPC}
\begin{equation}
    \xi (T ) = a \, \mathrm{exp} \left({\frac{b}{\sqrt{T/T_c-1}}}\right),
    \label{xiT}
\end{equation}
 where $b\simeq 1.5$ and $a$ is a non-universal coefficient. If the Fermi velocity remains constant within the considered temperature range, it is anticipated that the resistivity obeys the scaling law
\begin{equation}
    \rho (T) \sim \xi^{-1} \propto A \mathrm{exp} \left( -\frac{b}{\sqrt{T/T_c-1}} \right).
    \label{Rho}
\end{equation}
The resistivity vanishes at the transition temperature. It is important to note that in realistic systems, there is a complex interaction between the phase field and quasiparticle excitations. Nevertheless, the scaling behavior described by Eq.~(\ref{xiT}) remains applicable and only the value of $T_c$ is modified by this interaction. This assertion is based on the premise that the interplay between these factors does not significantly alter the underlying mechanisms governing the phase fluctuation and the BKT scaling behavior in Eq. (\ref{xiT}) is universal in the phase fluctuating regime of two-dimensional superconductors.
 
 Remarkably, as depicted in Figs.~\ref{Fig2}($\mathrm{a}$) and ~\ref{Fig2}($\mathrm{b}$) , Eq.~(\ref{Rho}) demonstrates a precise fit to the numerical data for both $s$- and $d_{x^2-y^2}$-wave pairing states. This accurate fitting yields an accurate estimation of the transition temperature {$T_c=0.089(2)$} for both pairing cases, implying that the correlation length of the phase field indeed dictates the electron scattering length.  We have also conducted a scaling analysis employing a formula proposed by Halperin and Nelson~\cite{Halperin1979JLTP}. Their formula accurately fits the numerical data for the $d_{x^2-y^2}$-wave pairing state but fails in fitting the results of $s$-wave pairing. However, in the $d_{x^2-y^2}$-wave pairing case, our scaling formula outperforms theirs in precision, offering a superior fit. Contrary to our findings, a recent study~\cite{Maska2020PRB} suggested that the phase fluctuations may lead to a linear-$T$ resistivity in the $s$-wave pairing state. This discrepancy likely results from the approximation and the relatively small system sizes used in their calculations. 
  
\begin{figure}[tbp]
    \includegraphics{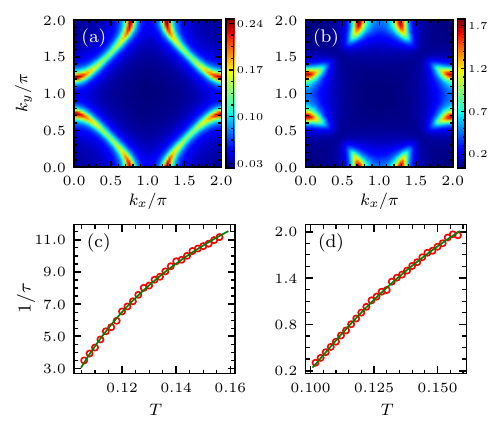}
    \caption{Momentum dependence of the imaginary self-energy for (a) $s$-wave and (b) $d$-wave pairing states at $T=0.11$, $\omega = 0.1$. The self-energy is calculated by taking an average of over $72$ samples with $L=70$ and $\eta = 0.008$. Panels (c) and (d) show the temperature dependency of the scattering rate of electrons around the Fermi surface along the ($\pi, 0$) direction of the Brillouin zone. The solid lines are fitting curves using the formula $1/\tau\sim \mathrm{exp}({-b}/{\sqrt{T\slash T_c-1}})$. }\label{Fig3}
\end{figure}

 In the above analysis, we find that the scattering lifetime of electrons follows the same temperature dependence of the correlation length of the fluctuating superconducting order parameters. To gain a deeper understanding of this observation, we calculate the self-energy of electrons using the Dyson equation
\begin{equation}
    \Sigma(k,\omega) = G^{-1}(k,\omega) - G_0^{-1}(k,\omega), 
\end{equation}
 where $G_0^{-1}(k,\omega)$ is the single-particle Green's function of free electrons, described by ${\cal H}_0$ without the superconducting pairing. The imaginary part of the self-energy, $\textrm{Im}\Sigma (k, \omega )$, measures the damping effect induced by the fluctuating phases. Its inverse equals the quasiparticle lifetime $\tau (\bs{q}, \omega)$. 
 
 Figs~\ref{Fig3}$\left(\mathrm{a}\right)$ and ~\ref{Fig3}$\left(\mathrm{b}\right)$ show the heat maps of the quasiparticle scattering rate, $1/\tau ({\bs q}, \omega)$, in the Brillouin zone for the $s$-wave and $d$-wave pairing states in a square lattice of $L=70$ at $\omega=0.1$, respectively. The result indicates that the quasiparticle scattering rates are the strongest around the Fermi surface close to ($\pi, 0$) and its equivalent points.  The momentum anisotropy in the scattering rate of $s$-wave pairing is presumably a consequence of the anisotropy of the spectral weight crossing the Fermi surface originating from the von Hove singularity in the band structure of electrons locating at momenta $(\pm\pi, 0)$ and $(0,\pm\pi)$.  For the $d_{x^2-y^2}$-wave pairing state, the anisotropy in the quasiparticle scattering rate is more pronounced. Particularly, the scattering rate almost vanishes along the nodal direction of the $d_{x^2-y^2}$-wave gap order parameter, indicating that the quasiparticles are more coherent along the nodal direction. This is consistent with the anisotropy of the $d_{x^2-y^2}$-wave pairing symmetry, 
 
 The lower panels of Fig.~\ref{Fig3} show the temperature dependence of the scattering rate $1/\tau ({\bs q}, \omega)$ at the points on the Fermi surface where the scattering rates are largest at $\omega = 0.1$. We average over eight equivalent points on the Fermi surface to obtain the data. The quasiparticle lifetime $\tau$ presumably relates to the scattering time. Therefore, we expect the corresponding scattering rate to share the same temperature dependence as the resistivity 
 \begin{equation}
     \tau^{-1} \sim  \mathrm{exp}\left(- \frac{b}{\sqrt{T\slash T_c-1}} \right) .
 \end{equation}
 This is indeed the case, as shown in Fig.~\ref{Fig3}. By fitting the data with the above expression, we estimate the critical temperatures to be $T_c=0.088(2)$ for both the $s$-wave and $d_{x^2-y^2}$-wave states with $L=70$. These values agree with the results of $T_c$ extracted from the resistivity data.    

\begin{figure}[tbp]
    \includegraphics{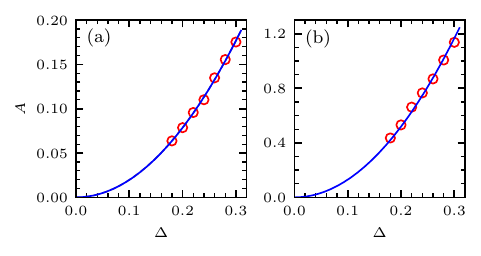}
    \caption{The resistivity coefficient $A$ defined in Eq.~(\ref{Rho}) versus the pairing gap amplitude $\Delta$ for (a) $s$-wave and (b) $d$-wave pairing states with $L=70$.}\label{Fig4}
\end{figure}

 Above the critical temperature, the leading-order contribution to the self-energy, $\Sigma(\mathbf{q},\omega)$, originates from the second-order perturbation of the fluctuating phase fields. Consequently, the primary contribution to the imaginary part of the self-energy is expected to exhibit a quadratic dependence on the gap amplitude $\Delta$ such that $A \propto \Delta^2$.
 
  To confirm this quadratic scaling behavior, we evaluated the resistivity as a function of temperature for various values of $\Delta$. Fig.~\ref{Fig4} illustrates the dependence of $A$ on $\Delta$, determined by fitting the numerical data to Eq.~(\ref{Rho}). Our findings reveal that the equation $A = c \Delta^2$, where $c$ is a coefficient independent of $\Delta$, accurately captures the numerical results. It is important to note that this quadratic term solely represents the lowest-order scattering process. The inclusion of higher-order corrections, such as a $\Delta^4$ term, further refines the fit. These results demonstrate a correlation between the resistivity coefficient $A$ and the amplitude of the fluctuating field, paving the way for a deeper understanding of the notable relationship between $A_1$ and $T_c$ observed in cuprate \cite{YuanJie2022Nat} and other unconventional superconductors \cite{JiangXY2023NP}, as reported in previous studies.
 
\subsection{Superfluid density}

 From the current-current correlation function and the expectation value of the diamagnetic current operator, we can also calculate the superfluid density in the superconducting phase: 
\begin{equation}
	\rho_s = -\langle K_x \rangle - \Lambda^{xx}(q_x=0, q_y\rightarrow 0, \omega = 0),
	\label{Supd}
\end{equation}
 where $K_x$ is the $x$-component of the diamagnetic current operator 
\begin{equation}
    K_x = -\frac{t}{N}\sum_{\bs{r}, \sigma} \left( c_{\bs{r},\sigma}^\dagger c_{\bs{r}+\bs{e}_x,\sigma} + c_{\bs{r}+\bs{e}_x,\sigma}^{\dagger} c_{\bs{r},\sigma} \right) .
\end{equation}

 The formation of the phase coherence determines the superconducting transition. For the two-dimensional XY model, the superfluid density exhibits a universal jump at the critical  point~\cite{Nelson1977PRL}, where the correlation length diverges. Exactly at the critical point,  the superfluid density scales linearly with $T_c$ 
\begin{equation}
    \rho_s\left(T_c, L\right) \simeq \frac{2}{\pi J_\theta}T_c. 
    \label{Eq:Tc_scaling}
\end{equation}

\begin{figure}[tbp]
    \includegraphics{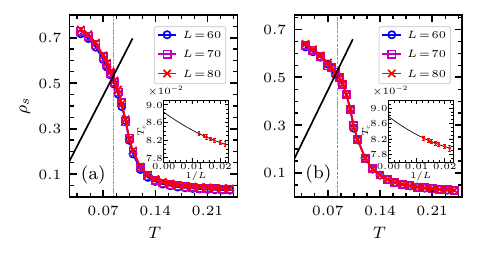}
    \caption{Temperature dependence of the superfluid density for both $s$-wave (left panel) and $d$-wave (right panel) pairing states. The black diagonal line represents the universal scaling law $\rho_s\left(T\right) \simeq {2T}/{\pi J_\theta}$, with its intersection with the superfluid density curve identifying the superconducting transition temperature $T_c$, denoted by the gray dashed line. The estimated $T_c$ from the superfluid density of $L=80$ is about $0.0835(4)$ for $s$-wave pairing and $0.0822(4)$ for $d_{x^2-y^2}$ wave pairing. Inserts in the two panels are extrapolations of the results with system sizes ranging from $46$ to $80$ by using quadratic polynomial, indicating  $T_c\simeq 0.088(2)$ in the thermodynamic limit.}\label{Fig5}
\end{figure}

 Fig.~\ref{Fig5} shows the superfluid density calculated in three different system sizes. With increasing temperature, the superfluid density exhibits a noticeable drop at a temperature close to the BKT transition temperature, $T_{KT} \simeq 0.893 J_{\theta}$, determined by the large-scale Monte-Carlo simulation ~\cite{Komura2012JPSJ, Nguyen2021AS}, indicating that the superconducting coherence is suppressed at high temperatures due to the phase fluctuation. 

 From the crossing point between the calculated superfluid density and the linear scaling formula (\ref{Eq:Tc_scaling}), we find 
 $T_c = 0.0835(4)$ and $T_c = 0.0822(4)$ when the system size $L=80$ for $s$-wave and $d$-wave pairing symmetries, respectively. Extrapolating the $T_c$ obtained from different system sizes, shown in the inserts of Fig.~\ref{Fig5}, we estimate $T_c$ to be $0.088(2)$ in the thermodynamic limit, close to the result of Monte Carlo $T_\mathrm{KT} = 0.0893$, independent of the pairing symmetry. It lends support to the scaling hypothesis of resistivity shown in Eq.~(\ref{Rho}). The finite superfluid density above $T_c$ is attributed to the finite-size effect.

\section{summary}

 In this work, we systematically investigate the electrical resistivity induced by the superconducting thermal phase fluctuation in two dimensions. Employing the Monte Carlo approach, we numerically solve the BdG Hamiltonian in the presence of classical pairing phase fluctuations governed by the ferromagnetic XY model. For both $s$-wave and $d_{x^2-y^2}$-wave pairings, the DC resistivity displays normal metallic behaviors, deviating from the conventional Fermi-liquid scaling properties. The DC resistivity induced by thermal phase fluctuations obeys the universal scaling dependence on the temperature $\rho\left(T\right)= A \mathrm{exp}({-b}/{\sqrt{T/T_c-1}})$, which is the same as critical scaling behavior of correlation length for BKT transition. The subsequent calculations indicate that the quasi-particle lifetime satisfies the same scaling relation, further substantiating the results of DC resistivity scaling behavior. Further analysis shows that the resistivity under superconducting fluctuations is related to the pairing amplitude, and the coefficient $A$ of the resistivity is a quadratic function of $\Delta$. The quadratic scaling relation potentially provides a direction to understand the interplay between unconventional superconductivity and anomalous dissipation in the normal states.  

Our results show that superconducting thermal phase fluctuations lead to novel transport properties in two dimensions. Since the temperature dependence of DC resistivity arising from pairing phase fluctuation unveiled here is not linear-T, our study cannot directly account for the scaling relation between the strange-metal dissipation and superconducting $T_c$ as observed in experiments. Nonetheless, as aforementioned, our results only apply to the thermal phase fluctuations in two dimensions, where the superconducting transition is a BKT transition. For the most unconventional superconductors, we should consider a three-dimensional system and incorporate the effects of inter-layer coupling. It is intriguing to study the DC resistivity arising from pairing phase fluctuation in a three-dimensional system, and particularly, investigate whether the strange-metal behavior could emerge or not. Moreover, notice that we only consider the electronic dissipation arising from pairing thermal phase fluctuations. Regarding realistic materials, various scattering mechanisms affect the transport properties above superconducting transition temperature $T_c$, including electron-electron scattering, electron-phonon scattering and disorder effect. Considering the cooperative effects on the electronic transport originating from pairing phase fluctuation and these various ingredients is also an interesting direction, which is left in our future study.

\textbf{Acknowledgments:} {We thank Kui Jin, Kun Jiang, Chui-Zhen Chen and Haiwen Liu for stimulating discussions. This work is supported by the National Natural Science Foundation of China, (Grant No.~12488201, No.~12322403, No.~12347107, No.~12488201, No.~11874095, and No.~11974396), the Strategic Priority Research Program of Chinese Academy of Sciences (Grants No.~XDB0500202, ~XDB33010100 and No.~XDB33020300), the National Key Research and Development Project of China (Grants No.~2022YFA1403900 and No.~2017YFA0302901) and the Youth Innovation Promotion Association of Chinese Academy of Sciences (Grant No.~2021004).}

\end{document}